\documentclass[twocolumn]{article}
\usepackage[utf8]{inputenc}
\usepackage[OT4]{fontenc}
\usepackage{amsmath}
\usepackage{graphicx}
\usepackage{multirow}

\title{Wandering in the state space.}
\author{Małgorzata J. Krawczyk\\
Faculty of Physics and Applied Computer Science,\\
AGH University of Science and Technology,\\
al. Mickiewicza 30, 30-059 Kraków, Poland\\
e-mail: gos@fatcat.ftj.agh.edu.pl
}
\date{\today}
\begin{document}
\maketitle

\begin{abstract}
We analyse the topology of the state space of two systems: \textit{i)} $N$ Ising spins $\pm 1$ with the antiferromagnetic interactions on a triangular lattice, with the condition of minimum of energy, \textit{ii)} a roundabout of three access roads and three exit roads, with up to $2$ cars on each road. The state space is represented by a network, and states - as nodes; two nodes are linked if an elementary process (spin flip or car shift) transforms the respective states one into another. Information is collected on the number of neighbours of states, what allows to distinguish classes and subclasses of states, and on the cluster structure of the state space. In the Ising systems, the clusters are characterized by anisotropy of the spin-spin correlation functions. In the case of a roundabout, the clusters differ by the number of empty or full roads. The method is general and it provides a basis for applications of the random walk theory.
\end{abstract}

\section{Introduction}
The aim of this paper is an introduction of a new method of analysis of a space of states of a given system. If the analysed system is not large, we are able to find all its possible states, and we can also separate states which demonstrate some property, e.g. minimum of energy. Subsequently we can investigate the topology of the state space, to indicate important properties of states. \\If we are able to define the way one state of a system can be transformed into another, we can think of a space of states as a network of states. Links in this network indicate possible transitions between states. If not all possible transitions are equally probable, the obtained network is weighted. The network of states can be analysed for identification of groups of nodes which are more densely connected to each other than to the remaining nodes in the system, i.e. the network of states can be divided into communities.

Existence of connections between states means that the diffusion or the random walk in such system can be analysed \cite{new}. Both those processes are interesting from the perspective of spreading information across the network. Such a phenomenon is important in many different areas of interest, to mention only sociology, economy or biology. Also communities are important from this point of view. The question is if we are able to indicate some characteristic properties of states which are classified to one community. If this is the case the time dependence of these properties can be analysed within the frames of the random walk theory.

The analysis is performed for two systems. The first one is the space of ground states of a triangular lattice with the antiferromagnetic Ising model. Its important feature is the frustration whose source lies in an even number of nearest neighbours of each spin and its two possible orientations. For an analysed system periodic boundary conditions are imposed. The number of possible states of the system strongly increases with the size of a lattice. Here we discuss the space of ground states, for lattices of two different numbers of nodes: $N=25$ and $N=36$. Triangular lattice was also one the systems analysed in our previous paper \cite{mk}. There we have shown that it is possible to pass by one-spin flip from one ground state into another. The number of states a given state can be transformed to is different for different states. Because of that the whole set of ground states can be divided into classes, where states which belong to a given class are connected by one-spin flip with the same number of other ground states. It was also shown \cite{mk} that in the case of a triangular lattice of size $25$ the space of ground states can be divided into three equally sized subsets, within which it is possible to move through all states changing an orientation of only one spin between two subsequent states. Such a situation indicates that for this system a kind of a constant of motion exists, which has three different values for those three subsets. The definition of this constant of motion is perhaps not unique. Here we present the results of the analysis of the correlation function within these subsets. What we also present is the calculation of the density of motives which are observed for the antiferromagnetic Ising model on triangular lattice \cite{shim}.

The second analysed system is the space of states of a roundabout. In general the number of roads, and particularly the number of vehicles which can occupy them, can be very large. Because of computational limitations we are able to analyse small systems only. It, however, does not change the generality of considerations. If the given number of access and exit roads is assumed, as well as a finite number of vehicles which can occupy each road, the system has a finite number of states. Here we analyse a system which consists of three access roads and three exit roads. The maximal number of vehicles on each road is limited to 2.\\
What we are interested in is the analysis of communities which possible states of the system can be divided into. Such a division indicates states which show similar properties, e.g. have the same number of fully occupied or empty roads.\\

The paper is organized as follows: in the next section the analysis of the triangular lattice is presented. We analyse correlation function for the lattice of $N=25$. The analysis of a density of characteristic motives for $N=25$ and $N=36$ is presented, and their connection with classes of states. The next section is devoted to analysis of the roundabout system. We analyse possible transitions between states, and their division into communities. The last section contains concluding remarks.

\section{Ising model on triangular lattice}
\begin{figure}[!hptb]
\begin{center}
\includegraphics[width=.3\textwidth, angle=270]{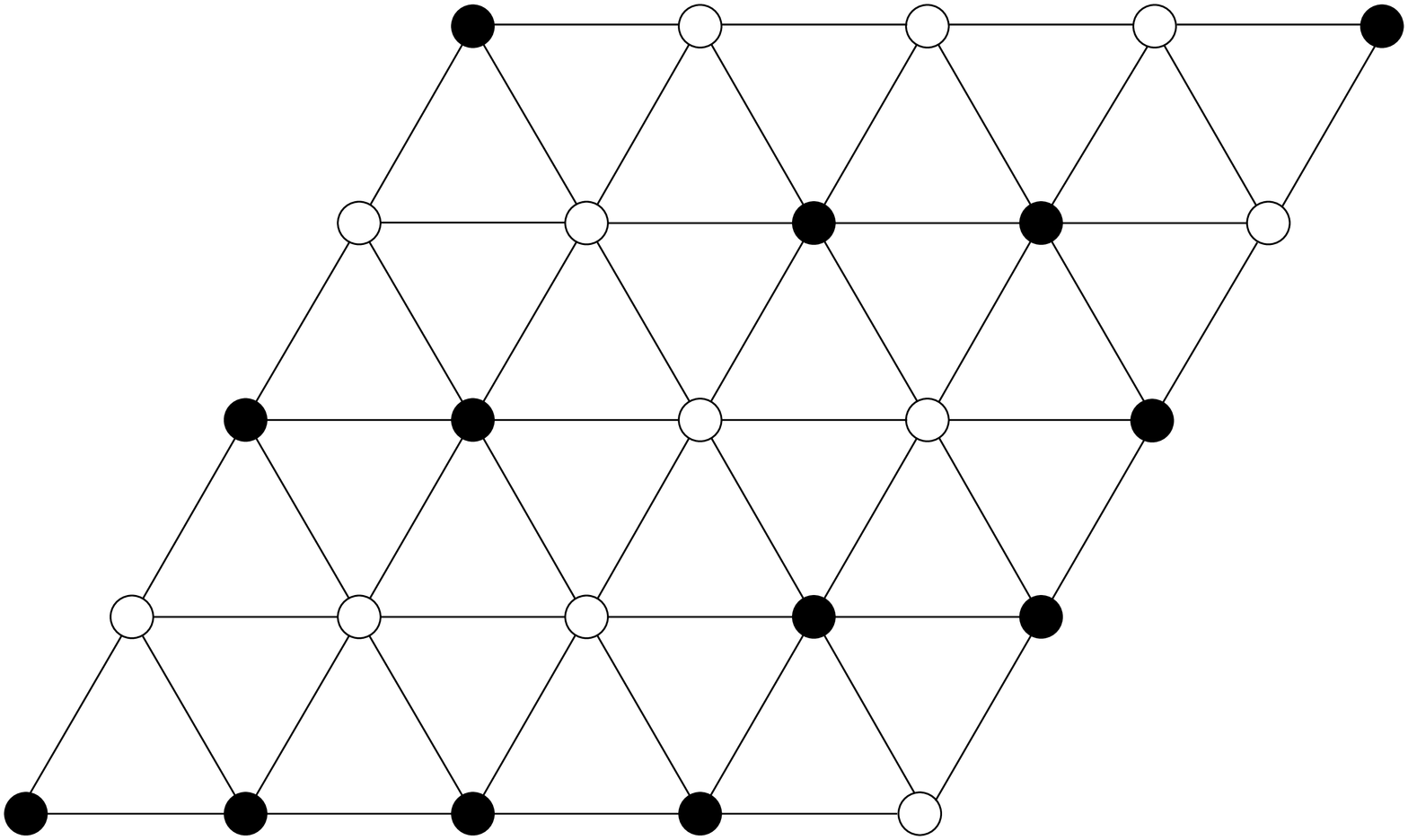}
\caption{Exemplary ground state of the triangular lattice of $N=25$.}
\label{fig0}
\end{center}
\end{figure}

The first analysed system is a triangular lattice the nodes of which are decorated by spins which fulfil the Ising model \cite{ising1, ising2, ising3}, i.e. two possible spin orientations $\sigma_i$ are allowed. Energy in the case of absence of an external magnetic field is given by the formula:
\[E=-J\sum\limits_{<i, j>}\sigma_i\sigma_j\]
where: $\sigma_i=\pm 1$. If an antiferromagnetic interaction is assumed, $J=-1$, this means that the preferred mutual spin orientation is antiparallel. In the above equation summing up goes via pairs of nearest neighbours.\\
The ground state energy for the lattice of $N=25$ is equal to $-25$ and for $N=36$ equals $-36$ which means that on average for each triangle there is one frustrated spin. The number of ground states in the first case is equal to $3630$ and in the second is $263\,640$. Exemplary ground state for the lattice of $N=25$ is shown in Fig.\ref{fig0}.

\subsection{Triangular lattice with N=25}
In our previous paper \cite{mk} it was shown that for the triangular lattice with $N=25$ nodes the whole set of ground states can be divided into three equally sized subsets which were obtained by means of the breadth-first search (BFS) algorithm \cite{bfs}. We start from a randomly chosen ground state, and all other ground states which can be achieved by one-spin flip from the selected one are marked. In the next step, configurations are marked transformable from those marked in the previous step. The procedure is repeated until all configurations are marked, or none can be selected. In the latter case, we check if another subset of ground states can be identified applying the same procedure for all states which were not marked so far. In the end, all $3630$ ground states were divided into three equally sized subsets $S$. This means an existence of a three-fold symmetry and a constant of motion with different values for three obtained subsets.

For each ground state the correlation function with nearest neighbours of each spin along three axis of the system is calculated. If we denote three directions of the triangular lattice as $x,\;y\;\text{and}\;z$, then:
\[
w_i=\langle \sigma_{-1i}\sigma_0+\sigma_0\sigma_{1i}\rangle_N,\quad\text{where}\;i\in[x, y, z]
\label{eq1}
\]
Obtained values are the same for all ground states, and are equal to $-0.6$, $0.2$, $-0.6$ (the order arbitrarily chosen). The difference between the three subsets is only in the order of values. This result indicates that in each subset a different axis is highlighted. So we observed an anisotropy of the correlation function.

\begin{center}
\begin{figure}[!hptb]
\hspace{-.5cm}
\begin{tabular}{ccc}
\begin{minipage}{.27\columnwidth}
\begin{center}
\includegraphics[width=1\textwidth, angle=270]{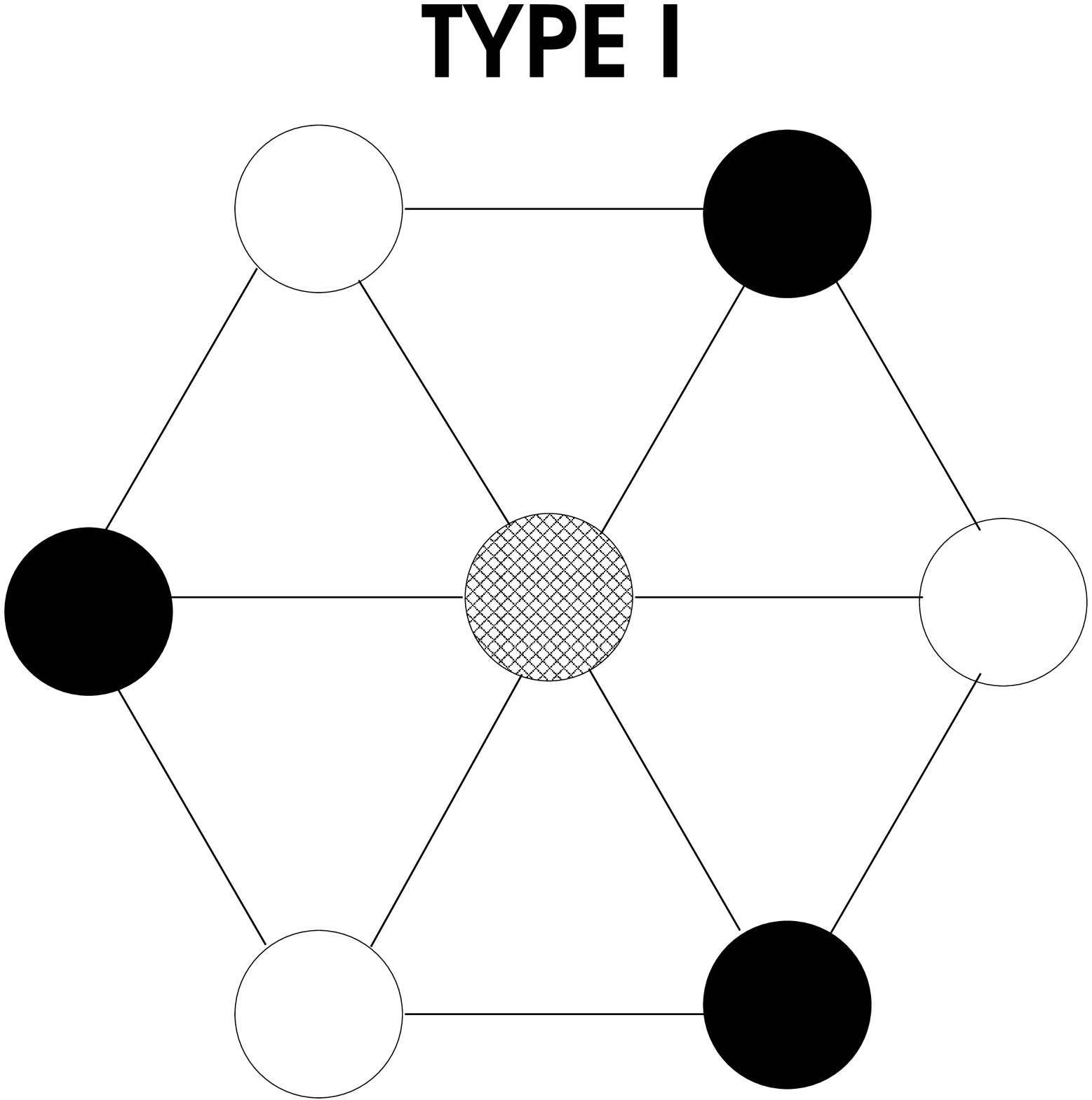}
\end{center}
\end{minipage}
&
\begin{minipage}{.27\columnwidth}
\begin{center}
\includegraphics[width=1\textwidth, angle=270]{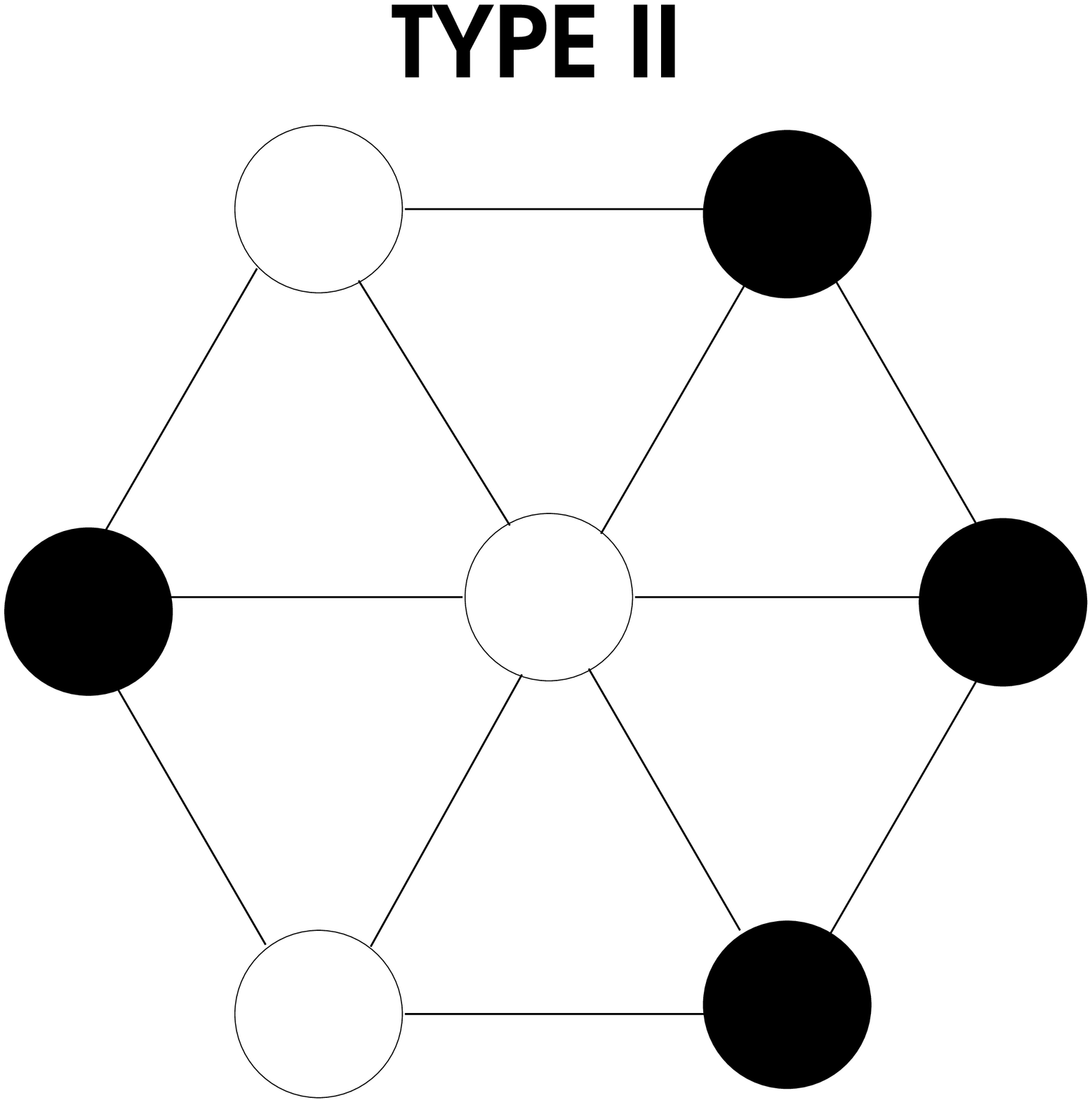}
\end{center}
\end{minipage}
&
\begin{minipage}{.27\columnwidth}
\begin{center}
\includegraphics[width=1\textwidth, angle=270]{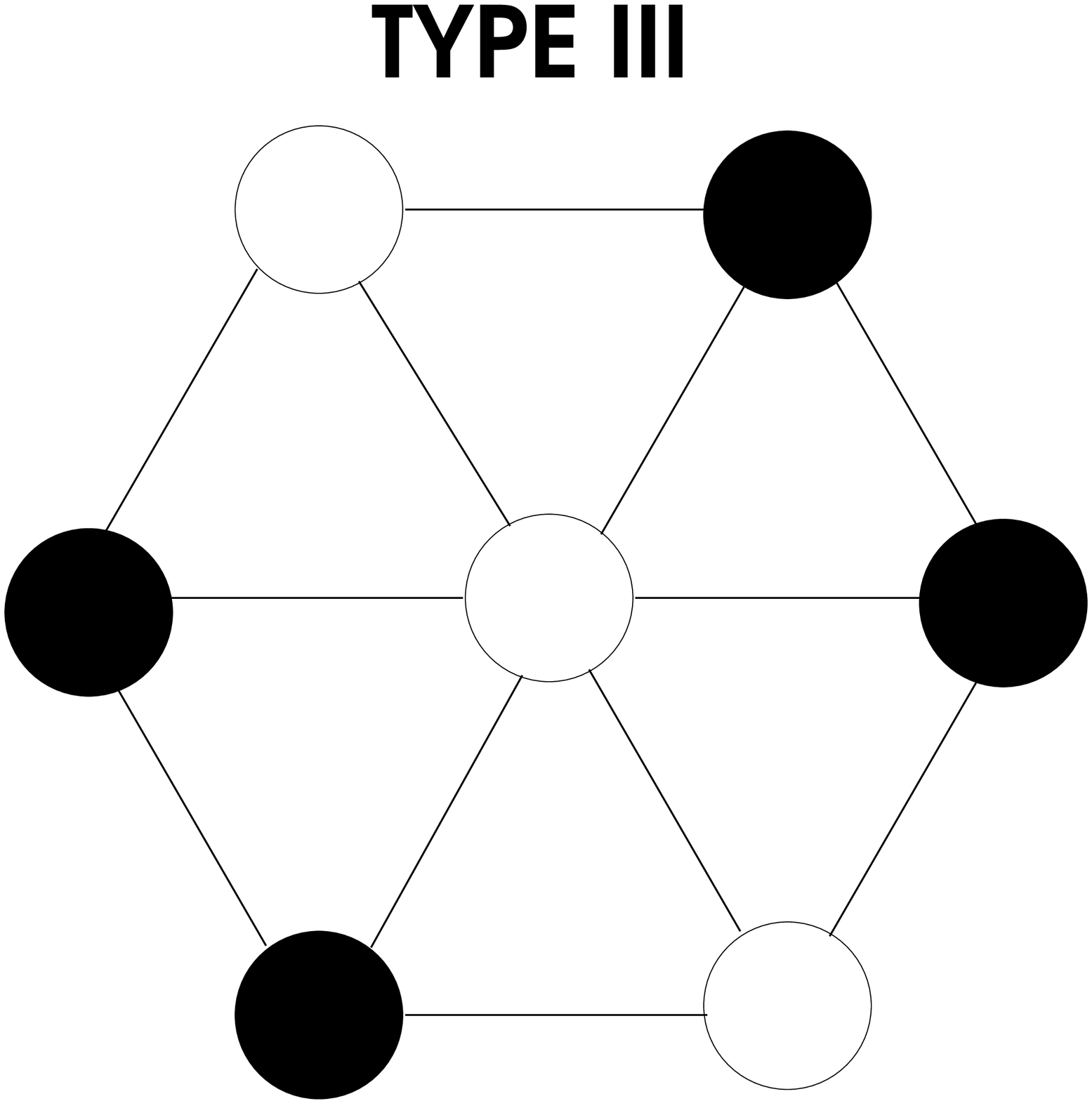}
\end{center}
\end{minipage}
\end{tabular}
\caption{Motives observed in ground state for a triangular lattice. White and black colour of nodes indicate two possible spin orientations (no difference which is ''down'' and which is ''up''), the grey colour means that spin can be either ''down'' or ''up''.}
\label{fig1}
\end{figure}
\end{center}

As it was shown in \cite{shim} in ground states of the Ising model on triangular lattice, characteristic motives are observed. Here we check if the density of these motives for given ground state is a quantity which can be used for indication of their specific properties, e.g. possibilities of its transformation into other ground states. Analysed motives are presented in Fig.\ref{fig1}. In \cite{shim} motives of \textsc{Type~II} and \textsc{III} were not distinguished. Here we show that their separation can be meaningful. White and black colour of nodes in the figure indicate two possible spin orientations, inversion of all spins is seen as the same motive. In the case of motive of \textsc{Type~I} the node in the middle (grey colour) can be with equal probability in one of two possible orientations. The reason is the fact that spin when three of his neighbours are ''up'' and three are ''down'' is frustrated. It means that for both its orientations the energy is exactly the same.

For each motive one should also consider its possible rotations around the central node. This, with combination of inversion, gives $2$ motives of \textsc{Type~I}, $12$ motives of \textsc{Type~II} and $6$ motives of \textsc{Type~III}. The way of spin arrangement in the ground state strongly depends on the size of a system, which is intuitively understandable. One can expect that if linear size of the lattice is a multiple of three, mindful of the two possible spin orientations, imposed periodic boundary conditions will allow for an alternating arrangement of ''down'' and ''up'' spins along one axis. 

Here we check if the anisotropy of the system manifests also in the density of different types of motives.

In the case of a lattice of size equal to $25$, not all spins are arranged in motives. The results are presented in Tab.\ref{tab1}.  The first column of this table covers a symbol of the class. The states belonging to a given class, indicated by the capital letter, can be transformed in a one-spin flip to the same number of different ground states. This number is denoted as $N_s$ (the second column in the table). Different subclasses, which are indicated by the digit following the class symbol, mean that states attainable from a given subclass belong to different classes (see \cite{mk}). The third column contains information about the number of states $N_c$ belonging to a given class. The three next columns include fractions of identified motives. Those values were calculated by inspection of the nearest neighbours of each node. If the arrangement of nodes fit to one of the wanted motives, its number is increased by $1$. In the end the obtained numbers of motives are divided by the number of nodes in the analysed network. 

The percentage of motives of \textsc{Type~I} is the same for all subclasses of given class. The same is true for the number of motives of \textsc{Type~II} in the case of class $C$ and $D$. For states belonging to class $E$, motives of this type are not observed. The highest variety of values is observed for motives of \textsc{Type~III}. In this case, the threefold symmetry of the system appears. For each subclass the whole amount of states is equally divided into groups inside which the triplets of equally oriented spins lay along one of the axes (see Fig.\ref{fig2}), and they are not observed along two remaining axes.\\
The same property is also observed within previously described subsets $S$. In each subset the motives of \textsc{Type~III} are oriented along one of the axes. This is an important property of these subsets.

\begin{center}
\begin{table}
\begin{center}
\begin{tabular}{|c||c|c|c|c|c|}
\hline
class&$N_s$&$N_c$&$f_I$&$f_{II}$&$f_{III}$\\\hline\hline
A1&3&300&0.12&0.44&0.32\\\hline
B1&4&600&0.16&0.40&0.28\\\hline
B2&4&300&0.16&0.48&0.20\\\hline
C1&6&150&0.24&0.32&0.28\\\hline
C2&6&300&0.24&0.32&0.24\\\hline
C3&6&300&0.24&0.32&0.20\\\hline
C4&6&600&0.24&0.32&0.24\\\hline
C5&6&300&0.24&0.32&0.20\\\hline
C6&6&150&0.24&0.32&0.24\\\hline
D1&8&300&0.32&0.16&0.20\\\hline
D2&8&300&0.32&0.16&0.20\\\hline
E1&10&30&0.40&0&0.20\\\hline
\end{tabular}
\end{center}
\caption{Fractions of identified motives for triangular lattice of 25 nodes. $N_s$~-~number of states given state can be transformed into, $N_c$~-~number of states belonging to a given class, and $f_i$~-~fraction of identified motives of given type.}
\label{tab1}
\end{table}
\end{center}

\subsection{Triangular lattice with N=36}
For lattice of $N=36$ nodes the obtained results are presented in Tab.\ref{tab2}. In this case, the ground states are divided into $20$ classes, and $409$ subclasses. The exact numbers of subclasses $N_{sub}$ which are obtained for different classes is presented in the second column of Tab.\ref{tab2}. In the next column of this table, the number of states $N_s$ to which a given state can be transformed in a one-spin flip is presented. As it can be seen this number ranges from $0$ to $24$. This means that for a triangular lattice of size $N=36$ isolated ground states are observed. Value $N_c$ (fourth column of the table) is the number of states which belong to a given class.\\Occurrence the isolated states implies that in this case it is impossible to pass through the whole space of ground states by one-spin flips. However, if we exclude them from the analysed space it is possible to pass through the whole space, state by state, by one-spin flips. 

In the case of the triangular lattice of $N=36$, $116$ possible combinations of the fractions of three analysed motives are observed. Values $N_t$ in Tab.\ref{tab2} indicate the number of different triplets of fractions of three motives $(f_I, f_{II}, f_{III})$ obtained for states which belong to a given class. However, as for the smaller lattice, the fraction of motives of \textsc{Type~I} $f_I$ is constant within given class of states (the last column of the table), and its value increases with increasing value of $N_s$. Obtained results indicate that in this case, states for which spins are arranged completely in accordance with the motive of \textsc{Type~I} are possible. This occurs for all $6$ ground states, which belong to the class $T$. Obtained value $f_I=0.67$ means that for two thirds of the nodes three neighbours are ''up'' and three are ''down'', and they are arranged alternately. The remaining one third are nodes whose orientation can be either ''up'' or ''down'' (marked in grey in Fig.\ref{fig1}). 
In turn, results obtained for states which belong to class $A$ indicate that in this case motives of this type are not observed.

\begin{center}
\begin{table}
\begin{center}
\begin{tabular}{|c||c|c|c|c|c|}
\hline
class&$N_{sub}$&$N_s$&$N_c$&$N_t$&$f_I$\\\hline\hline
A&1&0&186&3&0\\\hline
B&1&3&144&1&0.08\\\hline
C&2&4&540&2&0.11\\\hline
D&2&5&1296&2&0.14\\\hline
E&13&6&6072&7&0.17\\\hline
F&15&7&10800&6&0.19\\\hline
G&33&8&20052&9&0.22\\\hline
H&41&9&30816&10&0.25\\\hline
I&59&10&36864&11&0.28\\\hline
J&57&11&43344&12&0.31\\\hline
K&62&12&40428&13&0.33\\\hline
L&36&13&25200&9&0.36\\\hline
M&40&14&26208&11&0.39\\\hline
N&18&15&8400&5&0.42\\\hline
O&14&16&8244&5&0.44\\\hline
P&7&17&3312&4&0.47\\\hline
Q&4&18&1008&2&0.50\\\hline
R&2&19&576&2&0.53\\\hline
S&1&21&144&1&0.58\\\hline
T&1&24&6&1&0.67\\\hline
\end{tabular}
\end{center}
\caption{Classes identified in the triangular lattice of $36$ nodes. $N_{sub}$~denotes number of subclasses of given class, $N_s$~-~number of states given state can be transformed into, $N_c$~-~number of states belonging to a given class, $N_t$~number of different triplets $(f_I, f_{II}, f_{III})$, where $f_i$~is a fraction of motives of type~$i$, and $f_I$~-~fraction of motives of \textsc{Type I} for a given class.}
\label{tab2}
\end{table}
\end{center}

\begin{center}
\begin{figure}[!hptb]
\hspace{-.5cm}
\begin{tabular}{ccc}
\begin{minipage}{.27\columnwidth}
\begin{center}
\includegraphics[width=1\textwidth, angle=270]{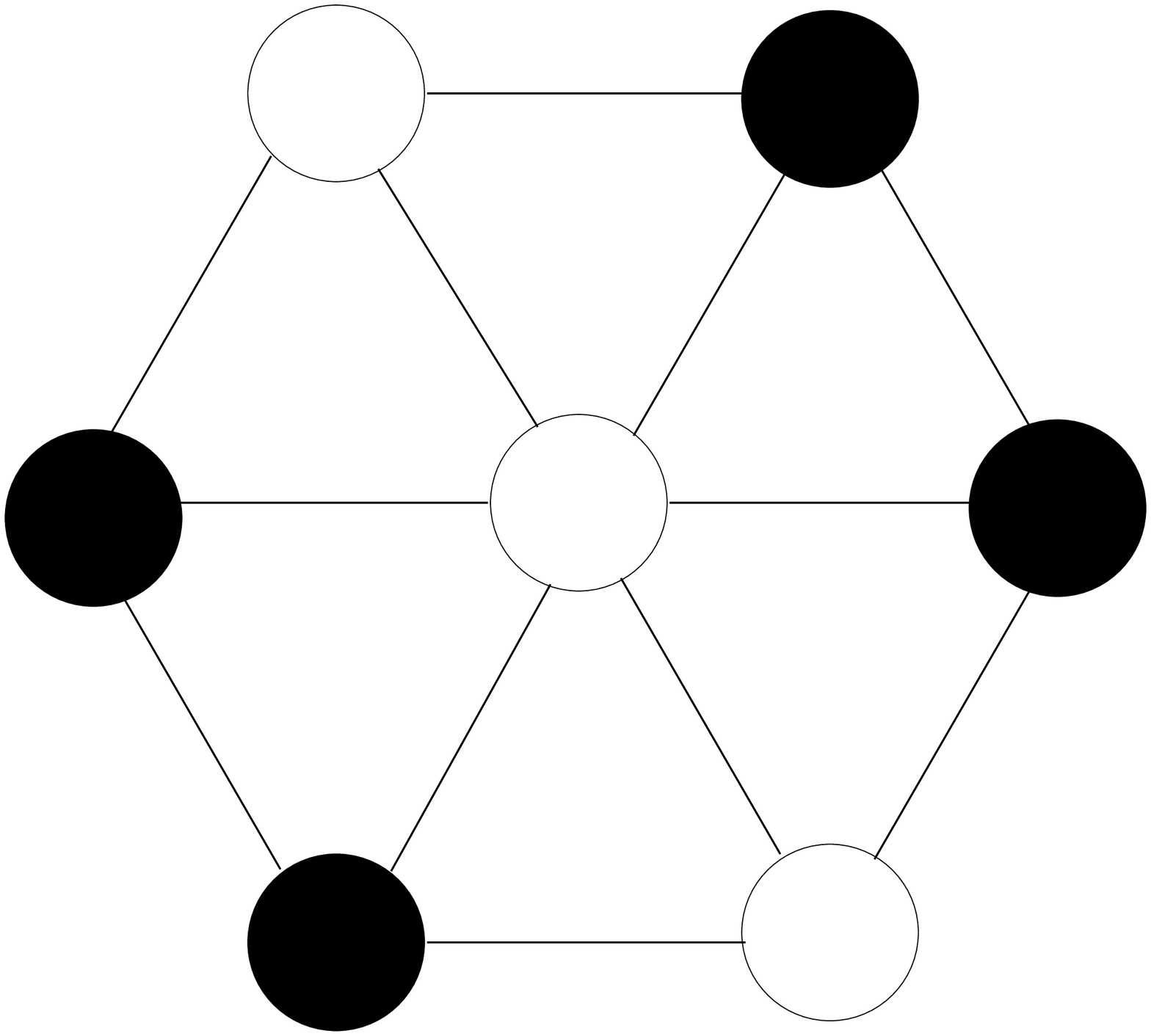}
\end{center}
\end{minipage}
&
\begin{minipage}{.27\columnwidth}
\begin{center}
\includegraphics[width=1\textwidth, angle=270]{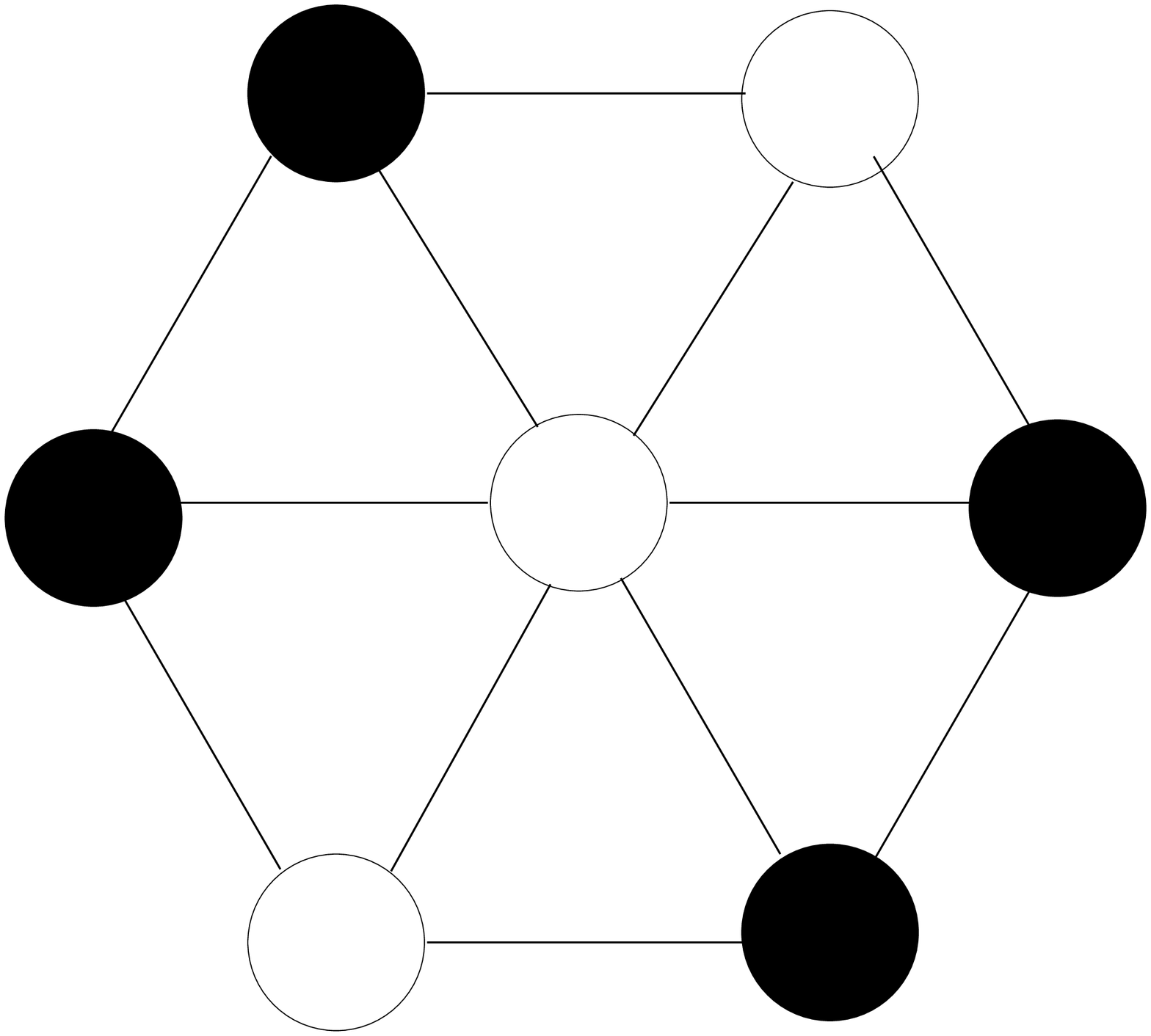}
\end{center}
\end{minipage}
&
\begin{minipage}{.27\columnwidth}
\begin{center}
\includegraphics[width=1\textwidth, angle=270]{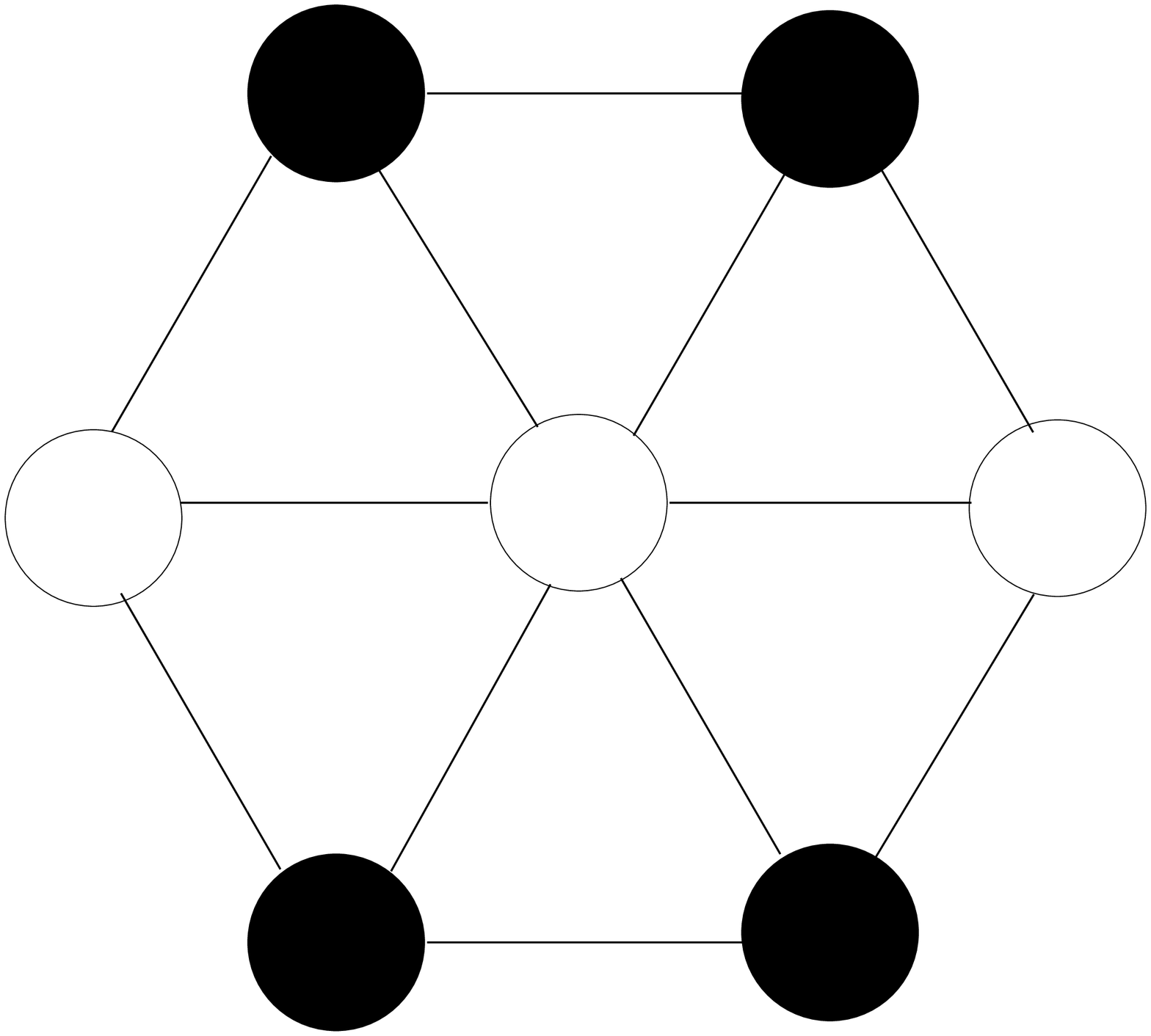}
\end{center}
\end{minipage}
\end{tabular}
\caption{Motives of \textsc{Type III} oriented along three axes of a system}
\label{fig2}
\end{figure}
\end{center}

Analysis of fractions of the motive of \textsc{Type~III} for lattice of $N=36$ is more complicated than for a smaller system. What we are interested in is if obtained subclasses can be divided into sets in which one axis is highlighted. This was the case for the lattice of $N=25$ where subsets of states were observed, where motives of \textsc{Type~III} were oriented only along a given axis (see Fig.\ref{fig2}).\\
Here 7 different possibilities are observed, each of which can be schematically presented as a triplet $T_{xyz}$ of values $(f_{IIIx}\,f_{IIIy}\,f_{IIIz})$, where the values relate to the three axes of the triangular lattice.\\
The first possible situation is that motives of \textsc{Type~III} are not detected, and in this case we have $(0\,0\,0)$. Other possibilities are: $(0\,0\,\gamma)$ - one non-zero value, $(0\,\beta\,\beta)$ - two equal non-zero values and $(0\,\beta\,\gamma)$ - two different non-zero values. Analogous triplets are observed for two remaining axes of the lattice, i.e. $(\alpha\,0\,0)$, $(0\,\beta\,0)$, etc. States for which non-zero fractions along each axis are also observed. In this case we have: $(\alpha\,\alpha\,\alpha)$ - three equal values, $(\alpha\,\alpha\,\gamma)$ - two equal values and $(\alpha\,\beta\,\gamma)$ - three different values. Also permutations which result from symmetry of the system are taken into account.

Tab.\ref{tab3} presents results of this analysis for the different classes of states. Here as a criterion of division the size of subclass $S_{sub}$ is used (the first row of the table). With the exception of class $A$ (analysed below) for states which belong to a given class obtained values $T_{xyz}$ are the same. The only difference is its order. The number of subclasses of a given size is denoted as $N_{S_{sub}}$. Subsequent rows contain the number of subclasses within which the scheme of obtained values fits to given triplet $T_{xyz}$. In these rows, the triplets with different permutations of the same numbers are added, i.e. the order of values in triplets is not taken into account.\\For states which form subclasses of size $1$ and $2$ all three values are always the same. In the case of subclasses of different sizes, only in some cases, motives oriented along one of the axes with their absence along remaining axes are observed.

An interesting situation is observed for states which belong to class $A$. Although this class is not divided into subclasses, different triplets $T_{xyz}$ were obtained for states which belong to this class. In the Tab.\ref{tab3} this class is presented in the last column (indicated by $\star$). For $6$ states belonging to this class, spins are arranged completely in accordance with the motive of \textsc{Type~III}, in two states along each of the axes of the lattice. For remaining states which belong to this class motives of \textsc{Type~II} and \textsc{III} are mixed: in $72$ cases in the ratio $1:2$, and in $108$ cases in the ratio $2:1$. In the first case a triplet $T_{xyz}$ of the type $(\alpha\,\beta\,\gamma)$ and in the second a triplet $T_{xyz}$ of the type $(0\,0\,\gamma)$ are obtained.

\begin{center}
\begin{table}
\begin{center}
\begin{tabular}{|c||c|c|c|c|c|c|}
\hline
$S_{sub}$&1&2&3&6&12&15\\\hline
$N_{S_{sub}}$&13&35&92&188&80&1$^{\star}$\\\hline
$N_{(0\,0\,0)}$&6&22&17&30&1&0\\\hline
$N_{(0\,0\,a)}$&0&0&21&28&16&1+1\\\hline
$N_{(0\,\beta\,\beta)}$&0&0&22&55&16&0\\\hline
$N_{(0\,\beta\,\gamma)}$&0&0&0&19&20&1\\\hline
$N_{(\alpha\,\alpha\,\alpha)}$&7&13&12&14&4&0\\\hline
$N_{(\alpha\,\alpha\,\gamma)}$&0&0&20&35&15&0\\\hline
$N_{(\alpha\,\beta\,\gamma)}$&0&0&0&7&8&0\\\hline
\end{tabular}
\end{center}
\caption{Analysis of motives of \textsc{Type~III} for classes identified in the triangular lattice of $36$ nodes. $S_{sub}$~denotes size of subclass, $N_{S_{sub}}$~-~number of subclasses of given size, $N_{(x\,y\,z)}$~-~number of states for which fraction of motives of \textsc{Type~III} along three axes of a lattice fulfils given scheme $T_{xyz}$ (an order in not important).}
\label{tab3}
\end{table}
\end{center}

\section{Roundabout system}
\begin{figure}[!hptb]
\begin{center}
\includegraphics[width=.3\textwidth, angle=270]{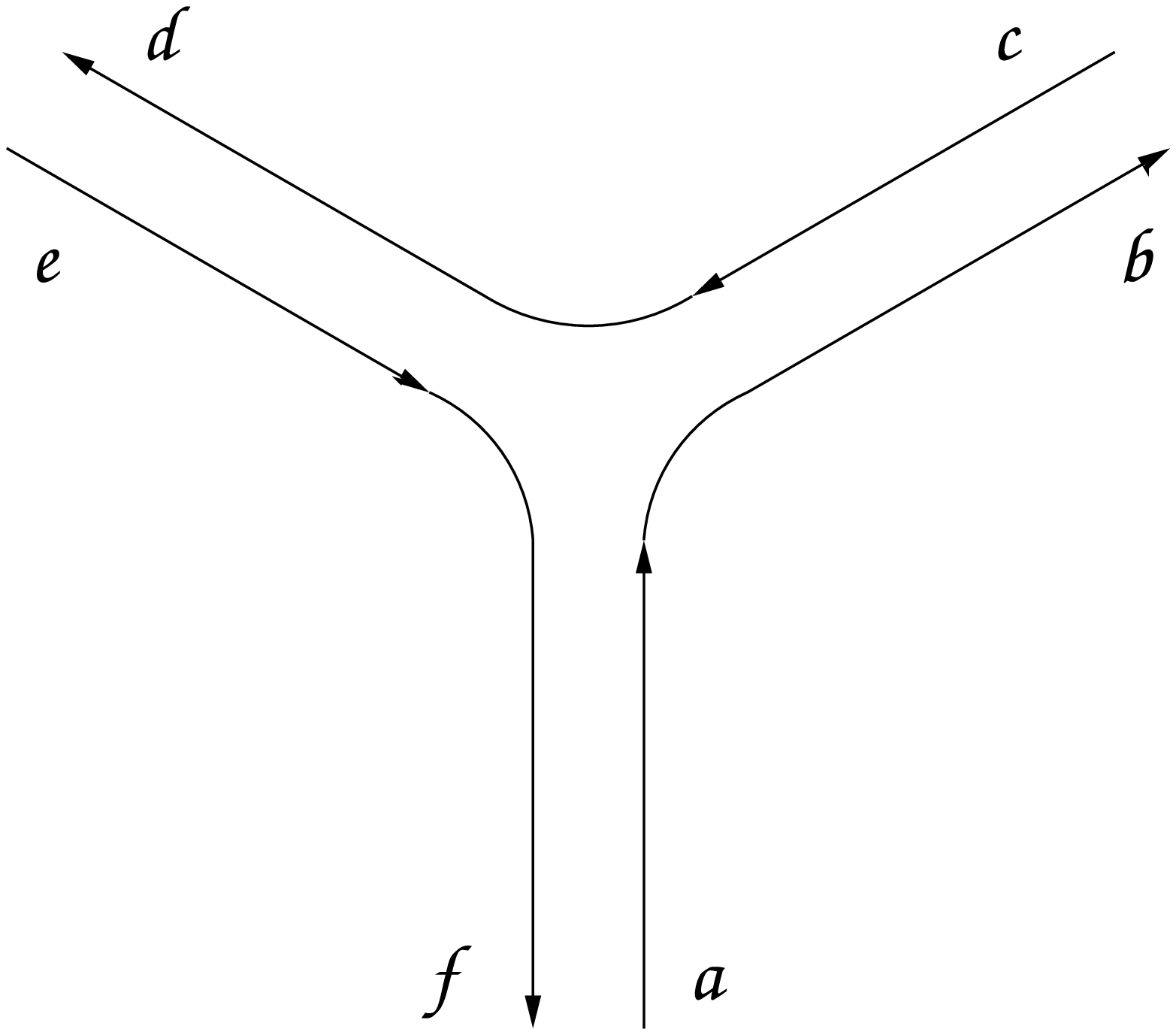}
\caption{Roundabout scheme with three access roads: $a\;c\;e$ and three exit roads: $b\;d\;f$.}
\label{fig3}
\end{center}
\end{figure}

The second analysed network is the state space of vehicles on a roundabout. Here we assume the existence of three access roads ($a\;c\;e$) and three exit roads ($b\;d\;f$) (Fig.\ref{fig3}). The maximal number of vehicles on each road is equal to $2$, so on each road two, one, or no cars is permitted. The state of the system is denoted as a sequence of six signs $(ab\;cd\;ef)$. In each pair the first digit refer to an access road, and the second - to an exit road. In other words, odd items identify access roads, while even - exit one. Each digit, as it was written before, can be in one of three possible states $0$, $1$ or $2$. For example, the state $(10\;21\;01)$ means that on the access road $a$ is one vehicle, on the access road $c$ two vehicles are present, one car is placed on exit roads $d$ and $f$, and no vehicles is on roads $b$ and $e$. Imposed limitations cause that the whole space of possible states contains $3^6=729$ states. 

If at most $2$ vehicles can be on each road at the same time, on an access road a new car can appear if the current occupation of this road is less then two. In the case when a state of any access road changes from $0$ to $1$ or from $1$ to $2$ the state of all remaining roads in the system remains unchanged. Also a decrease of the number of vehicles on an exit road does not influence the state of remaining roads. The movement of a vehicle from an access to an exit road is possible, if an access road is occupied at least by one car and an exit road is occupied by at most one car.

For each state a space of states attainable from a given one can be indicated. Each transition is connected with a change of the state of at least one road. The set of the transitions can be expressed as a directed connectivity matrix among possible states of a system. The obtained matrix is non-symmetric and weighted, as in some cases possible transitions are not equally probable (see below).

\begin{figure}[!hptb]
\begin{center}
\includegraphics[width=.3\textwidth, angle=270]{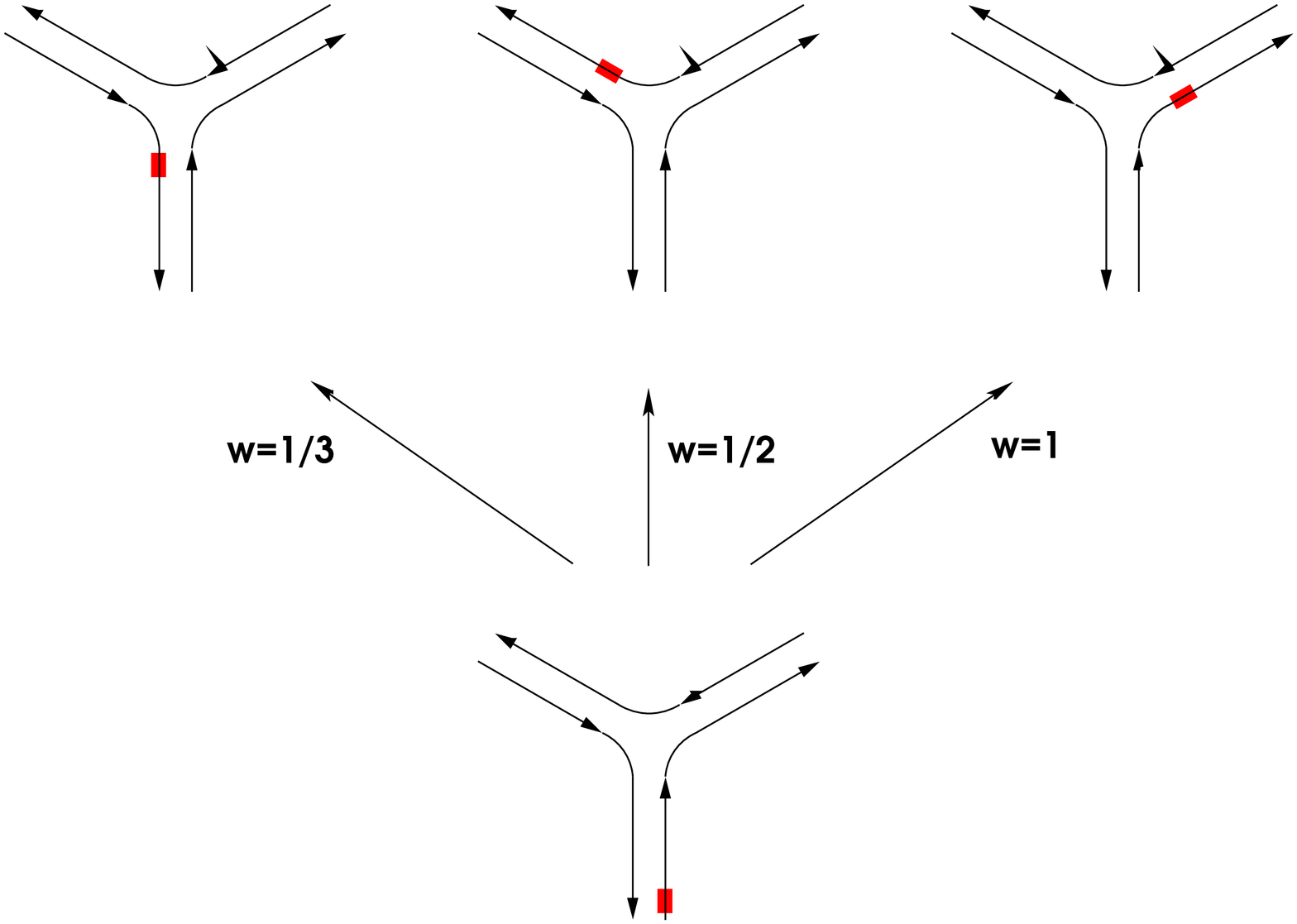}
\caption{State with one car can be transformed into three states, with different link weights which are dependent on the distance.}
\label{fig4}
\end{center}
\end{figure}

\begin{figure}[!hptb]
\begin{center}
\includegraphics[width=.3\textwidth, angle=270]{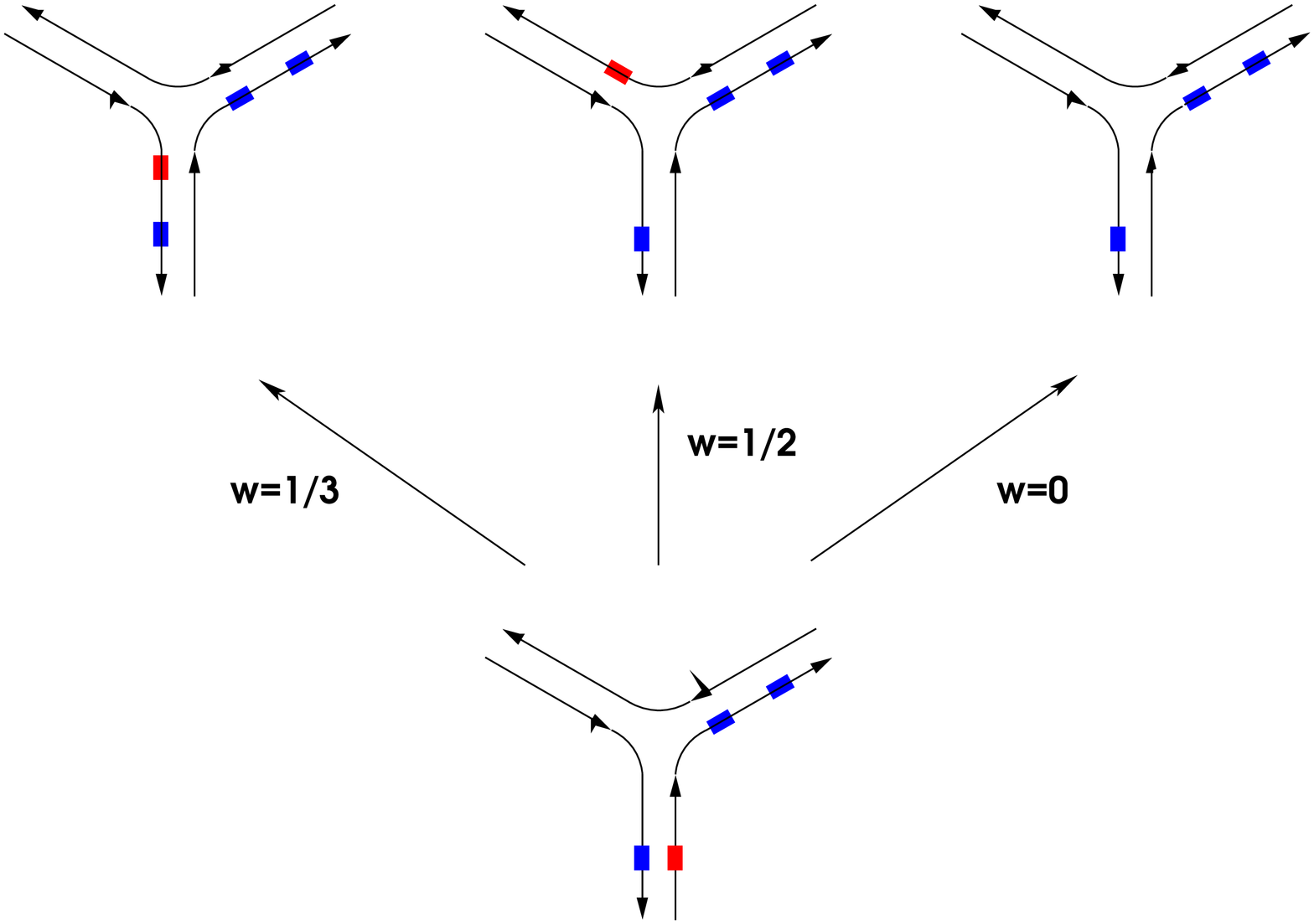}
\caption{Possible transformations depend on the number of cars.}
\label{fig5}
\end{center}
\end{figure}

When a roundabout is empty, a car can appear on any of the access roads, with equal probability. This means that the weight of a link from the state $(00\;00\;00)$  to each of the three states with one car on an access road $(10\;00\;00)$ or $(00\;10\;00)$ or $(00\;00\;10)$ is equal to $w=1$ . If a vehicle is on an access road (state with $1$ or $2$ on reads $a$, $c$ or $e$) a passage through the roundabout to one of the exit roads is possible. In this case however, the probability that a given exit road will be chosen depends on the availability of this road and a distance to cover (the distance on the roundabout passed by the car is treated by analogy to the electric resistance). An exit road is available if at most one vehicle is already there. Figs.\ref{fig4} and \ref{fig5} illustrate two of the possible situations. The first example shows possible transformation from the state $10\;00\;00$, i.e. from the state with one vehicle on one of the access roads (the situation is the same whichever access road is taken into consideration). The weight of a link between states equals $w=1$ for the closest exit road $(01\;00\;00)$, $w=1/2$ for the next one $(00\;01\;00)$, and $w=1/3$ for the last one $(00\;00\;01)$.\\The situation presented in Fig.\ref{fig5} shows the restriction of a number of cars on a given road. In our case if the number of vehicles on the road is equal to $2$ no more cars can enter. Because of that in the presented example one of the exit roads in unavailable, which cause that a link between state $(12\;00\;01)$ (on the bottom of the figure) and $(02\;00\;01)$ (top right) does not exist (weight of this link is equal to 0).\\
Similar considerations allow for specifying states which are attainable from each of the states of a system.

\subsection{Classes of states}
Likewise in a case of the triangular lattice, classes of states of the roundabout system were identified. Here, states are distributed among $13$ classes, the number of subclasses for each of them is presented in the second column of Tab.\ref{tab4}. The third column of this table covers information about the number of states ($N_s$) a state which belong to a given class can be transformed to, and the last column - the number of states belonging to a given class. 

\begin{center}
\begin{table}
\begin{center}
\begin{tabular}{|c||c|c|c|}
\hline
class&$N_{sub}$&$N_s$&$N_c$\\\hline\hline
A&1&3&2\\\hline
B&2&4&12\\\hline
C&6&5&57\\\hline
D&10&6&120\\\hline
E&7&7&141\\\hline
F&7&8&126\\\hline
G&6&9&103\\\hline
H&5&10&75\\\hline
I&4&11&45\\\hline
J&3&12&26\\\hline
K&2&13&15\\\hline
L&1&14&6\\\hline
M&1&15&1\\\hline
\end{tabular}
\end{center}
\caption{Classes identified in a roundabout system consisting of three access and three exit roads, with maximally two allowed vehicles on each of them. $N_{sub}$ - denotes number of subclasses of given class, $N_s$ - number of states given state can be transformed into, and $N_c$ - number of states belonging to a given class.}
\label{tab4}
\end{table}
\end{center}

The smallest number of possible transitions occurs for the state which consists just of zeros or twos. In the first case one car can occur on one access road, while in the second - one car can leave the system. The highest number of possibilities occurs when each road is occupied just by one vehicle. In this case a new can appear on one of the access roads, a car can pass from the access road to an exit road, or a car can leave the system (Tab.\ref{tab5}). The number of possibilities presented in the last column of the table is calculated taking into account different pairs of roads, i.e. in the first case from the state $(11\;11\;11)$ three equally probable transitions to states $(21\;11\;11)$, $(11\;21\;11)$ and $(11\;11\;21)$ are possible.

\begin{center}
\begin{table}
\begin{center}
\begin{tabular}{|c|c|c|}
\hline
initial&destination&number of\\
state&state&possibilities\\\hline\hline
\multirow{3}{*}{$(11\;11\;11)$}&$(21\;11\;11)$&3\\
&$(02\;11\;11)$&9\\
&$(10\;11\;11)$&3\\
\hline 
\end{tabular}
\end{center}
\caption{Possible transitions from the state $(11\;11\;11)$.}
\label{tab5}
\end{table}
\end{center}

\subsection{Communities}
States of the system were also divided into communities, which allow for classifying them into sets of states which demonstrate similar properties. Finding communities turns out to be possible for a symmetrised weighted connectivity matrix, with the use of the differential equation method \cite{dem1, dem2}. As a result, $8$ singular communities, $12$ of size of $7$, $6$ of size of $49$ and one community of size $343$ states are obtained. The group of singular communities form states completely full (two vehicles on each of the roads of the system), completely empty (no vehicles at all), and states for which $2(4)$ roads are fully occupied(empty) and $4(2)$ are empty(fully occupied) (Tab.\ref{tab6}). Interestingly enough, in all those states there is an equal number of fully occupied and empty access and exit roads, i.e. if only one access road is empty also only one exit road is empty.

The second group forms seven-element communities. Six of them have two fully occupied(empty) access roads and two fully occupied(empty) exit roads ($ab$ $cd$ or $ab$ $ef$ or $cd$ $ef$). Two other roads, one access and one exit, can be in any state other than $00$ and $22$. For remaining states, one access road and one exit road is full and ones access road and one exit road is empty, and as in the previous case two other roads can be in any state other than $00$ and $22$ (Tabs.\ref{tab7}, \ref{tab8}).

The next group consists of forty-nine-element communities. In this case for each state one access road and one exit road is fully occupied or empty. The state of the four remaining roads is any from the set of possible states with the exception of states which cause a state to belong to the previous group. Examples of states which belong to this group are: $(00\;12\;02)$ or $(21\;22\;10)$.\\All remaining states are classified together to one community.

\begin{center}
\begin{table}
\begin{center}
\begin{tabular}{|c|c|c|c|c|c|}
\hline
$a$&$b$&$c$&$d$&$e$&$f$\\\hline\hline
$0$&$0$&$0$&$0$&$0$&$0$\\\hline
$0$&$0$&$0$&$0$&$2$&$2$\\\hline
$0$&$0$&$2$&$2$&$0$&$0$\\\hline
$2$&$2$&$0$&$0$&$0$&$0$\\\hline
$0$&$0$&$2$&$2$&$2$&$2$\\\hline
$2$&$2$&$0$&$0$&$2$&$2$\\\hline
$2$&$2$&$2$&$2$&$0$&$0$\\\hline
$2$&$2$&$2$&$2$&$2$&$2$\\\hline
\end{tabular}
\end{center}
\caption{States which form singular communities.}
\label{tab6}
\end{table}
\end{center}

\begin{center}
\begin{table}
\begin{center}
\begin{tabular}{|c|c|c|c|c|c|}
\hline
$a$&$b$&$c$&$d$&$e$&$f$\\\hline\hline
$0$&$0$&$0$&$0$&$0$&$1$\\\hline
$0$&$0$&$0$&$0$&$0$&$2$\\\hline
$0$&$0$&$0$&$0$&$1$&$0$\\\hline
$0$&$0$&$0$&$0$&$2$&$0$\\\hline
$0$&$0$&$0$&$0$&$1$&$1$\\\hline
$0$&$0$&$0$&$0$&$1$&$2$\\\hline
$0$&$0$&$0$&$0$&$2$&$1$\\\hline
\end{tabular}
\end{center}
\caption{One of the communities of size $7$. The characteristic feature of this community is that two pairs of roads, here $a\;b$ and $c\;d$, are empty and values for the third pair can not be simultaneously equal either to $0$ or $2$. Two similar communities were obtained in which pairs of non-$0$ values are for roads $a\;b$ and $c\;d$, e.g. $00\;11\;00$. The next three communities cover states with $2$s in place of former $0$s on analogous pairs of roads, e.g. $22\;22\;11$.}
\label{tab7}
\end{table}
\end{center}

\begin{center}
\begin{table}
\begin{center}
\begin{tabular}{|c|c|c|c|c|c|}
\hline
$a$&$b$&$c$&$d$&$e$&$f$\\\hline\hline
$0$&$0$&$2$&$2$&$0$&$1$\\\hline
$0$&$0$&$2$&$2$&$0$&$2$\\\hline
$0$&$0$&$2$&$2$&$1$&$0$\\\hline
$0$&$0$&$2$&$2$&$2$&$0$\\\hline
$0$&$0$&$2$&$2$&$1$&$1$\\\hline
$0$&$0$&$2$&$2$&$1$&$2$\\\hline
$0$&$0$&$2$&$2$&$2$&$1$\\\hline
\end{tabular}
\end{center}
\caption{One of the six similar communities of size $7$. The characteristic feature of this community is that for one pair of roads, here $a\;b$, both values are equal to $0$, for one pair, here $c\;d$, are equal to $2$. Remaining communities of this type arise by swapping those pairs, and their shift (e.g. $22\;00\;01$ or $22\;01\;00$ etc.).}
\label{tab8}
\end{table}
\end{center}

A question appears; states belonging to which class are classified to the same community. The answer is as follows. Two states of the class $A$ were classified as singular communities. Six remaining communities of this type include states belonging to the same subclass of class $C$. All states of class $B$ were classified to one of the communities of size $7$. Remaining states of these communities belong to class $C$, $D$, $E$ or $F$. Whereas in $6$ out of $12$ cases states belong to classes from $B$ to $E$, and in the $6$ remaining, from $C$ to $F$. In communities of size $49$, states belonging to classes from a range $C-I$ were placed. The biggest community cover states from classes $C$ to $M$. The exact numbers of states of a given class classified to a given community are presented in Tab.\ref{tab9}. As it can be seen there in the case of communities of size $7$ two equally sized subsets can be indicated. In the first one, states belong to classes from $B$ to $E$, in the second - from $C$ to $F$. All communities of size $49$ cover exactly the same number of states of a given class, and their subclasses.

\begin{center}
\begin{table}
\begin{center}
\begin{tabular}{|c||c|c|c|c|c|c|}
\hline
$N_{cs}$&\multicolumn{2}{c|}{$1$}&\multicolumn{2}{c|}{$7$}&$49$&$343$\\\hline\hline
$N_m$&2&6&6&6&6&1\\\hline
$N_A$&1&-&-&-&-&-\\\hline
$N_B$&-&-&2&-&-&-\\\hline
$N_C$&-&1&1&1&6&3\\\hline
$N_D$&-&-&3&3&8&36\\\hline
$N_E$&-&-&1&2&12&51\\\hline
$N_F$&-&-&-&1&10&60\\\hline
$N_G$&-&-&-&-&8&55\\\hline
$N_H$&-&-&-&-&4&51\\\hline
$N_I$&-&-&-&-&1&39\\\hline
$N_J$&-&-&-&-&-&26\\\hline
$N_K$&-&-&-&-&-&15\\\hline
$N_L$&-&-&-&-&-&6\\\hline
$N_M$&-&-&-&-&-&1\\\hline
\end{tabular}
\end{center}
\caption{Indication of classes for states in a given community. $N_{cs}$ denotes the number of states classified to given community, $N_m$ number of communities with given scheme of classes within, and $N_{class}$, where $class$ in $[A-M]$, denotes the number of states within a community belonging to a given class (exact description in the text).}
\label{tab9}
\end{table}
\end{center}

Particularly interesting are isolated states, as they can be seen as ''special'' states. They can perform a crucial role in traffic, either as states which allow for reduction of traffic jams, or on the contrary can be a source of it. In some cases the solution can be found in the same system. If a given road is fully occupied the roundabout can be left through an empty exit road, even if such resolution is connected with a higher cost. In some situations it is worth taking a longer way than to get in a traffic jam.

\section{Discussion}
In this paper we have analysed the space of states of two systems: the Ising model on a finite triangular lattice and the roundabout with access roads and exit roads. In both cases the space forms a network of states. The method of our analysis is general and it can be applied to any other systems, where states can be described with discrete numbers. The most important limitation of the method is that the number of states strongly increases with the system size. However, a new insight can be reached also for small and moderate systems.

For the triangular lattice of size $N=25$ a new analysis of the correlation functions and density of characteristic motifs has been added to our previous results presented in \cite{mk}. From the density of different motifs, an anisotropy of the correlation function is found within three separate subsets (communities) of states. The number of subsets is a consequence of the three-fold symmetry of the triangular lattice. More general, the symmetry of the state network reflects the symmetry of the investigated system. Within the communities, this symmetry can be broken. In the case of the Ising system of $25$ nodes, what is broken is the rotational symmetry of the triangular lattice, as we observe anisotropic spin-spin correlation function within the communities. On the contrary, the up-down symmetry of the Ising model remains preserved within the communities, as one can reverse all spins flipping them one by one without an increase of energy.

The lattice of $N=36$ nodes is analysed here for the first time. In this case we found some isolated states, i.e. communities which contain one state only. In these states, a flip of any spin leads to an increase of energy of the system. Having these states excluded, we found that there is a path through all remaining ground states on which subsequent moves are one-spin flips. Analysis of the density of motives also in this case allows to indicate common properties of states in the state network. Also, the number of neighbours of states in the state network was used for
the indication of classes of the ground states of the system. Further, this classification is enriched to subclasses by taking into account the classes of the neighbours. The differentiation procedure is closed in two steps. This is an important difference between our networks of states and disordered networks; in the latter, the number of subclasses could reach to the number of nodes.

Our second example - a roundabout - is entirely different from the first one, and this difference can be seen as an argument on the universality of the method. Here also the division of states into classes and subclasses has been made. On the contrary to the previous example the state space is connected, then a community here is a set of states connected mutually more densely than to states from other community \cite{new1,fort}. The results of the method of differential equations \cite{dem1,dem2} indicate that the communities found in the roundabout differ in the number of empty or full pairs of roads. Therefore, possible transitions between the communities of the roundabout may be important if we look in the broader perspective. Formation of traffic jams on one of the roundabouts can be a source of communication problems in a wider area. Monitoring of the whole communication network allows for an indication of areas with already existing traffic jams or on which the high probability of their formation occurs; this in turn allows for suggesting drivers to use detours. This application will be discussed in a separate work.

Collecting data on the topology of the state space are important and should be useful in constructing the Master Equations for the time evolution of the probability of a given system. In particular, the diffusion on networks is known \cite{new} to depend directly on the network topology. With the data on classes, subclasses and communities, it should be possible to determine the time dependence of observables, which are characteristic for the communities of a given systems. Further applications of the data could include an identification of bottlenecks and calculation of exit times from particular communities to the rest of the space network.

{\bf Acknowledgement:} 
The author is grateful to Krzysztof~Kułakowski for critical reading of the manuscript and helpful discussions. The research is partially supported within the FP7 project SOCIONICAL, No. 231288.

\end{document}